# Time Dilation and the Equivalence of Inertial Frames

Chandru Iyer
*Plant Head (Retd), Lydall Performance Materials, Gurugram, Haryana, India*
Email: chandru_i@yahoo.com

G.M. Prabhu
*Dept. of Computer Science, Iowa State University, Ames, IA 50011*
Email: prabhu@iastate.edu

## Abstract

It is well known that simultaneity within an inertial frame is defined in relativity theory by a convention or definition. This definition leads to different simultaneities across inertial frames and the well-known principle of relativity of simultaneity. The lack of a universal present implies the existence of past, present and future as a collection of events on a four dimensional manifold or continuum wherein three dimensions are space like and one dimension is time like. However, such a continuum precludes the possibility of evolution of future from the present as all events exist 'forever' so to speak on the continuum with the tenses past, present and future merely being perceptions of different inertial frames. Such a far-reaching ontological concept, created by a mere convention, is yet to gain full acceptance. In this paper, we present arguments in favour of an absolute present, which means simultaneous events are simultaneous in all inertial frames, and subscribe to evolution of future from the present.

**Keywords:** Special relativity, conventionality of simultaneity, one-way speed of light, time dilation, synchronization

## 1. Introduction

The Michelson-Morley experiment [10] concluded that the round trip speed of light is constant as observed by all inertial frames in relative motion with respect to each other. It was realised that the one-way speed of light is indeterminable experimentally as it would require synchronization of spatially separated clocks within an inertial frame. In an isotropic frame a system of spatially separated clocks can be generated by slow separation of clocks. With the round trip speed of light being constant and independent of inertial frames, such a situation can be sustained only with moving clocks running slow. This immediately meant that a moving inertial frame would be anisotropic as clocks moving at the same speed in different directions within a moving frame will have different speeds with respect to the stationary frame.

The question of isotropy and anisotropy has been discussed in the context of two inertial frames in relative motion [7]. It is shown that if an inertial frame perceives an isotropic environment, then an inertial frame moving with respect to the isotropic inertial frame will perceive an anisotropic environment in the context of the mass and momentum distribution within a finite universe. It was also observed that if we assume the existence of an infinite universe, then it could be visualised that all inertial frames are isotropic and equivalent.

The definition of simultaneity as a result of a convention has been discussed in detail in [6]. It is pertinent to quote from the first and last paragraph of this comprehensive work.

“In his first paper on the special theory of relativity, Einstein indicated that the question of whether or not two spatially separated events were simultaneous did not necessarily have a definite answer, but instead depended on the adoption of a convention for its resolution. Some later writers have argued that Einstein’s choice of a convention is, in fact, the only possible choice within the framework of special relativistic physics, while others have maintained that alternative choices, although perhaps less convenient, are indeed possible.”

“The debate about conventionality of simultaneity seems far from settled, although some proponents on both sides of the argument might disagree with that statement.”

Reichenbach [11] has emphasised the conventionality of the Einstein-Poincare synchronization convention [2] and recommended that a parameter $\epsilon$ be introduced for every inertial frame and the one-way speed of light be $\frac{c}{1-\epsilon}$ in the forward direction and $\frac{c}{1+\epsilon}$ in the reverse direction. This maintains the round trip speed of light to be $c$, in agreement with the Michelson-Morley experiment [10].

Rugh and Zinkernagel discuss the concept of time and its physical basis [12]. The physical basis of the time concept is necessarily related to physical processes, which could conceivably take place among the material constituents within an object or an inertial frame. Therefore, time cannot be defined by mathematics and altered by choice of coordinates but only through ‘scale setting’ physical processes and their durations. In this context, Rugh and Zinkernagel [12] discuss (i) black body radiation, (ii) half time (decay) of unstable particles and (iii) well-defined radiation frequencies of transition between atomic states. Thus, the concept of time is fundamental to our perception and modelling and understanding of physical processes. In Section 2, we describe how anisotropy and the slow running of physical processes influence distant synchronization of clocks.

It has been pointed out in [4] that the adoption of the defined synchronization convention leads to reversal in time order of contiguous events. Further in [5], the experimental observation of relativistic Doppler effects are shown to be a geometric mean of the two classical Doppler effects – one assuming the speed of light is $c$ in the inertial frame co-moving with source and the other assuming the speed of light is $c$ in the inertial frame co-moving with the receptor.

This paper is organized as follows. Section 2 describes a process of slow separation of clocks and a derivation of the time dilation function that maintains symmetry between inertial frames. Sections 3 and 4 describe a thought experiment to synchronize three equivalent clocks in relative motion. Section 5 contains a discussion of clocks in cosmic time scales. Section 6 consists of the conclusion of the existence of a unique isotropic inertial frame.

## 2. Slow separation of clocks

The crux of the problem is evinced in the unique synchronization adopted by each inertial frame and the symmetry achieved in that scheme. When we consider two inertial frames (K and K′), any particular clock of K′ as observed by clocks of K appears to run slow and similarly any particular clock of K as observed by clocks of K′ appear to run slow. The rate at which this slowdown is observed is given by the expression $\sqrt{1 - v^2/c^2}$ that is defined as $1/\gamma$. The synchronization can be accomplished either by using light propagation, setting the one way speed of light as $c$ or by slow separation of clocks within the inertial frame. Both these methodologies give raise to the same synchronization [1, 6, 8, 9].

Let us consider an isotropic inertial frame, with a premise that clocks moving at speed $v$ with respect to this isotropic inertial frame run at rate $f(v)$ times the rate at which the clocks in this isotropic inertial frame run. If there were a set of identical clocks, all at a given location in the isotropic inertial frame, these can be synchronized manually. When we move these clocks apart (except one) in different directions, during the time interval of motion, the clocks will run slow and thus will lose their synchronization with respect to the clock that was left unmoved at the original location. However, all the moving clocks will maintain their synchronicity as they are moving at the same speed with respect to the isotropic frame, although in different directions. The same logic will not be applicable in the moving reference frame as it is anisotropic.

Further, the time dilation function needs to be an even function of speed in the isotropic reference frame as the time dilation with velocity $+v$ and $-v$ is identical and hence the time dilation is an even function, that is, $f(v) = f(-v)$; additionally the time dilation function has to be continuous with respect to $v$ at $v = 0$ and therefore, the time dilation function is a function of $v^2$, that is $f(v)$ does not have components of odd powers of speed in its Taylor series expansion in an isotropic frame. When we separate the clocks very slowly at a speed of $\Delta v$, in the limit $\Delta v \rightarrow 0$, the time dilation is negligible as $(\Delta v)^2$ is pronouncedly negligible compared to $\Delta v$. Consider an anisotropic frame K′, moving at a speed $u$ with respect to an isotropic inertial frame K. Let us say the same procedure of slow separation of clocks is executed in this anisotropic frame. A clock separated by a distance of $s$ by slow separation, will have an asynchronization of $sf'(u+v)_{at\ v=0} = sf'(u)$, as explained below (see Fig.1.).

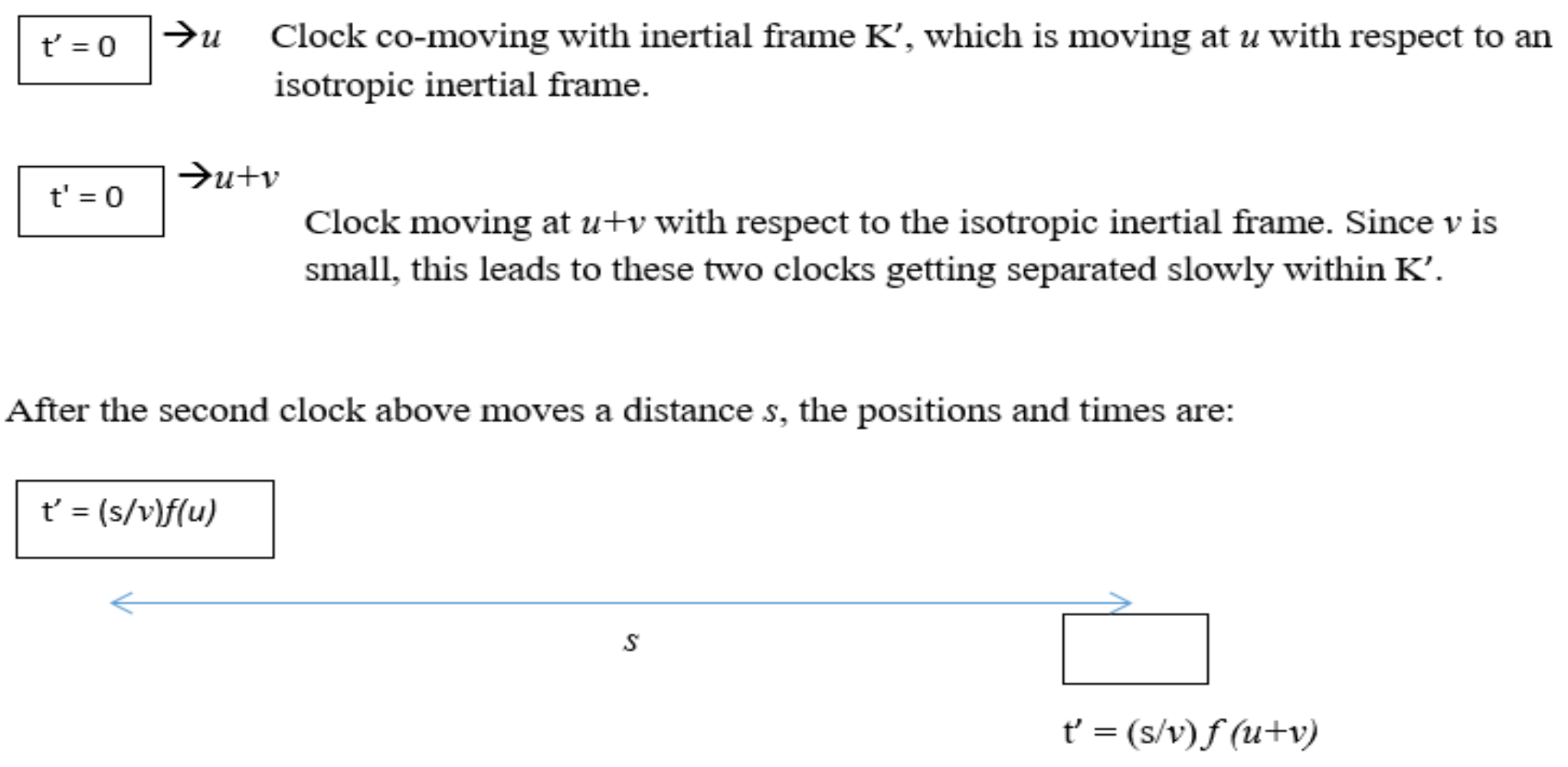

**Fig. 1. Two clocks getting slowly separated in inertial frame K′ moving at *u* with respect to an isotropic inertial frame**

---

Asynchronization = Limit $v \rightarrow 0$ $[(s/v)\, f(u+v) - (s/v)\, f(u)] = s\, f'(u)$

For an isotropic inertial frame, $u$ = 0 and $f'(0) = 0$, ($f(u)$ being an even function). Thus, there will be no asynchronization by this process for an isotropic inertial frame.

The clocks in K′ and K that are both synchronized by the process of slow separation of clocks in their respective inertial frames will be as shown in Fig. 2.

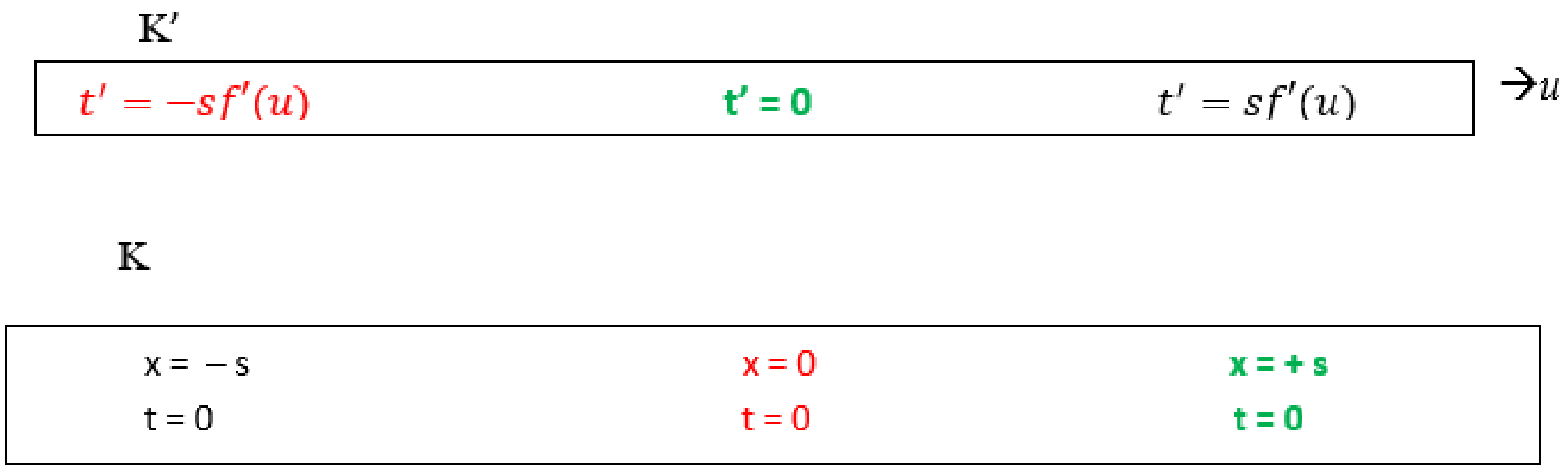

**Fig. 2. Asynchronization between K and K′ after both inertial frames follow slow separation process**

After an elapsed time of $s/u$, due to the movement of objects in K', the clocks marked in red and green colors will meet each other. Due to time dilation, the moving clock will run up a time of $f(u)(\frac{s}{u})$ and therefore the time shown by the clocks will be as depicted in Fig. 3.

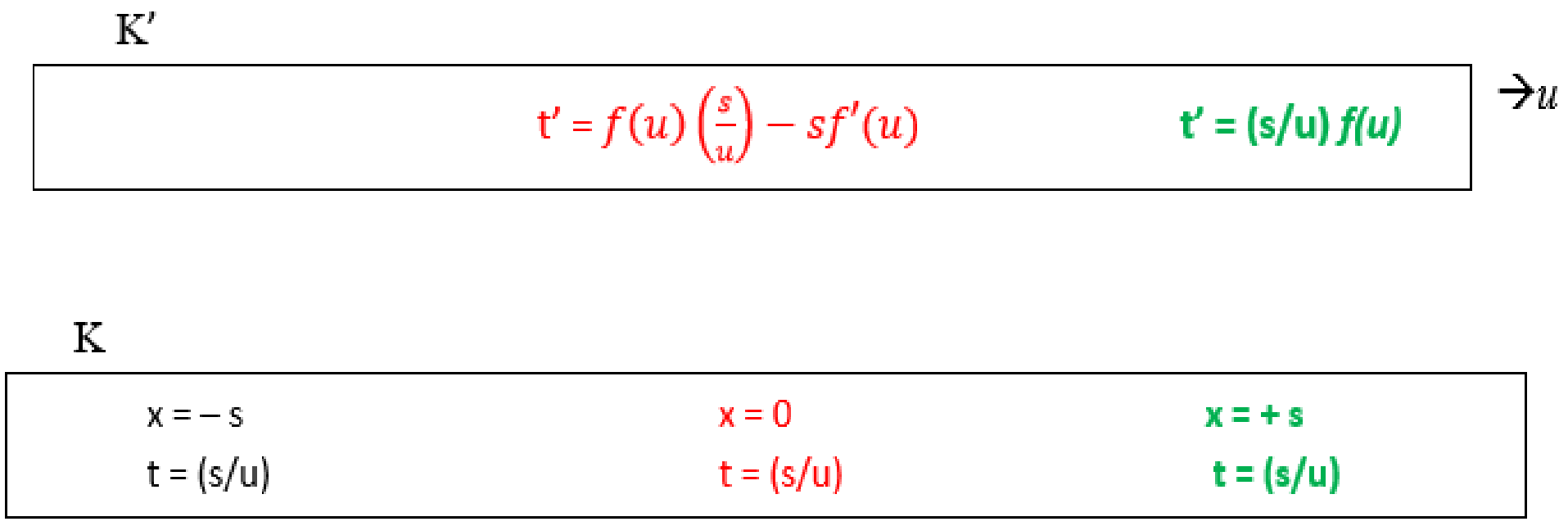

**Fig. 3. Space-Time Points at instant (s/u) in the isotropic inertial Frame K**

When the anisotropic frame is unaware of its own anisotropicity, it can consider that the 'stationary' clock has run up a time of $(s/u)$ in a time interval

$$f(u)\left(\frac{s}{u}\right) - sf'(u)$$

When this ratio $(s/u)/[f(u)\left(\frac{s}{u}\right) - sf'(u)]$ (that is the ratio of expressions marked in **red**), is equal to the time dilation of moving clocks as observed by the stationary inertial frame, that is $f(u)$, (the ratio of expressions marked in **green**), there is an apparent symmetry in that slow separation of clocks adopted by both the stationary and moving inertial frame give a perception that time dilation is mutual and identical as observed by the two inertial frames about each other.

Thus, when the function $f(u)$ satisfies the differential equation

$$f(u) = (s/u)/[f(u)\left(\frac{s}{u}\right) - sf'(u)] \tag{1}$$

there is an apparent perception of mutual slow running of clocks as observed by both the frames, with the synchronization of spatially separated clocks achieved by slow separation method in both the inertial frames. After cancelling *s* and simplifying, equation (1) becomes

$$f(u) = 1/[f(u) - uf'(u)] \tag{2}$$

Solving this differential equation, we obtain

$$f(u) = \sqrt{1 - u^2/c^2}$$

as detailed below.

Setting *f (u)*, the rate at which moving clocks run slow = *y*, we have two alternative solutions shown in Table 1.

| | |
|---|---|
| $y = 1/[\, y - u(dy/du)]$ | $y = 1/[\, y - u(dy/du)]$ |
| $y^2 - yu(dy/du) = 1$ | $y^2 - yu(dy/du) = 1$ |
| Separating variables, we have | Separating variables, we have |
| $-y\, dy/(1 - y^2) = du/u$ | $y\, dy/(y^2 - 1) = du/u$ |
| letting $z = 1 - y^2$ , we have | letting $z = y^2 - 1$ , we have |
| $(1/2)\, dz/z = du/u$ | $(1/2)\, dz/z = du/u$ |
| Integrating, we get | Integrating, we get |
| $(1/2) \log z = \log u + \log (1/c)$ | $(1/2) \log z = \log u + \log a$ |
| where $\log (1/c)$ is the integration constant | where $\log a$ is the integration constant |
| Thus $z = u^2/c^2$ | Thus $z = u^2a^2$ |
| or $1 - y^2 = u^2/c^2$ | or $y^2 - 1 = u^2a^2$ |
| or $f(u) = \sqrt{1 - u^2/c^2}$ | or $f(u) = \sqrt{1 + u^2a^2}$ |

**Table 1. Two solutions for *f (u)* starting with the same differential equation**

The solution in the right column would mean faster running of moving clocks. In accordance with observed time dilation, we select the solution in the left column and obtain

$$f(u) = \sqrt{1 - u^2/c^2} \tag{3}$$

where $c$ emanates from the integration constant and is set to the maximum one-way speed in all directions in the isotropic frame.

The key points in the analysis are that

- We assumed that moving clocks run at a different rate depending on their speeds.
- We used the same process of slow separation of clocks to synchronize spatially separated clocks in both the stationary and moving inertial frames.
- We derived the time dilation function so as to create an apparent symmetry; that is, under the synchronization achieved by slow separation of clocks, both the inertial frames observed that the clocks in the other frame are time dilated in the same manner.

Based on this analysis one may be tempted to conclude that since the observation of both the frames are identical about the other frame, the two frames are equivalent by mathematical symmetry. However, if the clocks in both the inertial frames run identically, the slow separation should produce identical synchronization in both the frames. The fact that the two inertial frames achieved different synchronizations by the process of slow separation of clocks implies that the clocks in both the inertial frames did not run identically.

It is interesting to note that the constancy of the average round trip light speed emerges as a corollary of this symmetry in asynchronicity. All these features, namely, relative movement, length contraction, asynchronization, and time dilation work in tandem to yield the Lorentz Transformation. According to the Lorentz Transformation, every inertial frame is isotropic internally, its clocks are running perfectly, its distant clocks are synchronized (identically by either the slow separation of clocks or by assuming the one-way speed of light to be constant) and every inertial frame is anisotropic when observed from another inertial frame with attendant asynchronization of distant clocks and slow running of clocks. However, this apparent symmetry has to be interpreted in the context of the inherent asymmetry between two inertial frames. *If both the inertial frames were isotropic and their clocks ticked identically, there will be no disagreement between the observers in the two inertial frames on distant clock synchronization.*

## 3. Synchronizing equivalent clocks across inertial frames

In view of the above apparent symmetry that emanated from an inherent asymmetry, it is interesting to investigate which of the clocks in K or K′ ran slow if at all. Special Relativity Theory (SRT) proposes that both the frames are equivalent in all respects implying that none of the clocks ran slow, that is, all the clocks both in K and K′ ticked identically [2]. However, as explained in the previous section, if the clocks in K and K′ ticked identically and if K and K′ were both isotropic, both the frames will achieve identical distant clock synchronization, there will be absolute simultaneity, and there will be no observation of slow running of moving clocks. Since this is not the case one is constrained to conclude that clocks in K and K′ did not tick identically.

However, a particular clock in inertial frame K will meet a particular clock in inertial frame K′ only once and hence they are not in a position to compare themselves for a second time to ascertain whether any one of them is running slow compared to the other. This aspect can be overcome when three clocks are in relative motion. In [3] we have considered three clocks *k*, *m* and *n* in relative motion, that is to say the pairs *(k, m), (m, n)* and *(n, k)* are all in relative motion such that three pairs of meetings of *(k, m), (k, n),* and *(n, m)* take place in that temporal order and designated as events $E_1$, $E_2$, and $E_3$ respectively. It is possible to envisage that at $E_1$ clocks *k* and *m* synchronize their time readings and at event $E_2$, clock *n* sets its time to be same as that of clock *k*. Now both clocks *m* and *n* have set themselves against *k* and therefore when *m* and *n* meet at event $E_3$, provided that all the three clocks *k, m,* and *n* run equivalently, *m* and *n* should display the same time at event $E_3$. However, if moving clocks run slow, then *m* and *n* will be running at different rates and hence will not show the same time at event $E_3$. So if clocks *m* and *n* show the same time at $E_3$, then one can say that all clocks run equivalently and there is no time dilation. On the other hand if time dilation exists, that is, if moving clocks run slow, then clocks *m* and *n* will not show the same time at event $E_3$. Such an observation at event $E_3$ of clocks *m* and *n* not showing the same time will validate time dilation but will lead to the conclusion that moving clocks indeed run slow and are not equivalent to stationary clocks.

Time dilation means that moving clocks run slow. The above attempt to synchronize three clocks assumes that the three clocks that are in uniform motion run equivalently. The prediction of special theory of relativity that such a synchronization will fail due to time dilation (due to slow running of moving clocks) means that the three clocks did not run equivalently and therefore, the three inertial frames co-moving with the three clocks are not equivalent.

## 4. Thought experiment to synchronize three equivalent clocks in relative motion

We illustrate the concepts described in Section 3 by a thought experiment framed through a numerical example. In this example we use the units *year* and *light-year* for *time* and *distance* respectively.

Let us say that an inertial observer (A) observes another inertial observer (B) passing him and they both synchronize their clocks to show a time of 0.0 (event $E_1$). After some time when the clock with observer A is showing a time of 0.1, observer A observes another inertial observer D, passing him (in the same direction as that of B) and D sets his time to 0.1 as shown by the clock of A (event $E_2$). At a later time, when D crosses B, they both observe that the clock of D is showing a time of 0.28 and the clock of B is showing a time of 0.32 (event $E_3$). Based on the three events, a scientist may conclude that the three clocks of A, B and D did not run at the same rate; or clocks in inertial frames in relative motion do not run identically, as otherwise at the meeting of B and D, both clocks would have shown the same time.

In the above numerical example, it may appear that we have used one of the three inertial frames associated with the three clocks (in this case that of clock A) as a preferred inertial frame. However, we may not perform any adjustment of clocks (do not make time = 0 at the first meeting of A and B and do not make time = 0.1 at the meeting of A and D), and instead note down the difference in time shown by the clocks at their respective meetings as $(\Delta)_{ab}$, $(\Delta)_{da}$, and $(\Delta)_{bd}$.

Let $(\Delta)_{ab}$ = (time shown by clock A – time shown by clock B) when they meet at event $E_1$.

Let $(\Delta)_{da}$ = (time shown by clock D – time shown by clock A) when they meet at event $E_2$.

Let $(\Delta)_{bd}$ = (time shown by clock B – time shown by clock D) when they meet at event $E_3$.

Then the sum of these time differences $(\Delta)_{ab} + (\Delta)_{bd} + (\Delta)_{da}$ should be equal to zero if the three clocks A, B and D run identically. In the above example, this sum will always remain 0.04 years. When we examine this sum in this perspective, we do not use any frame as a preferred frame but can see that the three clocks did not run equivalently. The events and observations as described in the previous paragraph sets $(\Delta)_{ab} = 0$ ; $(\Delta)_{da} = 0$ and $\Delta_{bd}$ is observed to be 0.04 years. But if we just note down the difference in time at the meetings of A, B and A, D without setting the differences at these events as zero and observe the time difference at the meeting of B, D at event $E_3$, then also the sum $(\Delta)_{ab} + (\Delta)_{bd} + (\Delta)_{da}$ will remain = 0.04 years. Thus, we have a general observation without appearing to have used the frame associated with clock A, or any other inertial frame, as a preferred frame. This shows that independent of observers associated with any frame of reference, clocks in relative uniform motion do not run equivalently, a conclusion that is not frame dependent.

We provide below an explanation of these results by Special Relativity Theory (from the inertial frame co-moving with clock A).

$E_1$: Clocks A and B meet at the same space-time point and synchronize their clocks as 0.

$E_2$: Clock A and clock D meet. At this event clock A shows a time of $t = 0.1$ ; clock D sets its clock as $= 0.1$.

At this instant, in the inertial frame co-moving with clock A, the time is 0.1 and clock B is at a distance of $0.1 * 0.6 = 0.06$ light-years from clock A.

Clock B is running slow as observed by the inertial frame co-moving with clock A by a factor of $\sqrt{1-0.6^2} = 0.80$ ; therefore, clock B will show a time $0.1 * 0.8 = 0.08$.

The time taken for clock D to reach clock B (as observed by observers co-moving with clock A) is $\left[\frac{0.06}{0.8-0.6}\right] = 0.3$.

Therefore the instant in inertial frame A, when clocks B and D meet is $0.1 + 0.3 = 0.4$.
At this instant clock B will show a time $0.4 * 0.8 = 0.32$.

In the time elapsed between $E_2$ and $E_3$, that is, 0.3 years, clock D which is running slow at a rate $\sqrt{1-0.8^2} = 0.60$, will show an elapsed time of $0.3 * 0.6 = 0.18$ years. This time when added to the initial time of 0.1 will yield 0.28 years as shown by clock D at $E_3$.

Thus, at event $E_3$ clock B will show a time of 0.32 years and clock D will show a time of 0.28 years, giving $\Delta_{bd}$ to be 0.04 years, as analyzed by observers co-moving with clock A.

Similar explanations by SRT can be constructed from the perspective of each of the three inertial frames associated (co-moving) with the three clocks or any other external inertial frame. In every such consideration, the times shown by clocks B and D will come out to be the same i.e., 0.32 years and 0.28 years respectively. This is as it should be, under special relativity, because the presence of clocks B and D at the same space-time point together is a *here and now* observation that will be observed identically, in terms of the times clocked by B and D, by observers in all inertial frames. The explanation by SRT implicitly assumes at every step that clocks in inertial frames do not run equivalently in whichever inertial frames the observers are situated. The algebraic derivation for the general case is given in the appendix in [3].

Table 2 describes the coordinates of the three events $E_1$, $E_2$ and $E_3$ as observed by the three inertial frames co-moving with the three clocks A, B and D. Space-time diagrams from the three inertial frames are depicted in the Appendix section.

| | Frame Co-moving With Clock A | Frame Co-moving With Clock B | Frame Co-moving With Clock D |
|---|---|---|---|
| $E_1$ | 0, 0 [2] | 0, 0 [2] | 0.13333,-0.066667 [1,4] |
| $E_2$ | 0, 0.1 [2] | -0.075, 0.125 [3] | 0, 0.1 [2] |
| $E_3$ | 0.24, 0.4 [3] | 0, 0.32 [2] | 0, 0.28 [2] |

**Table 2**: **Event co-ordinates in the three frames expressed as (x, t)**

The numbers 1, 2, 3, 4 in the upper right corner of each cell correspond to the observations/computations as described in the Note section below.

The Minkowski distances, as evaluated by all the three inertial frames are as follows:

$E_1$ to $E_2$ = 0.1
$E_2$ to $E_3$ = 0.18, and
$E_1$ to $E_3$ = 0.32

Path traced by A and D from $E_1$ to $E_3$ via $E_2$ is $E_1 E_2 + E_2 E_3 = 0.28$
Path traced by B from $E_1$ to $E_3$ that is the distance $E_1 E_3 = 0.32$

In order that B and D meet at $E_3$, either A or D or both have to 'move' slower along the continuum and cover a distance $E_1 E_2 + E_2 E_3 = 0.28$ in the same interval that B covered the distance $E_1 E_3$ (= 0.32).

Note:

1. Co-ordinates for $E_1$ in frame D are calculated by the Lorentz transformations and not actually observed by D.
2. These co-ordinates are actual observations with synchronization performed at $E_1$ and $E_2$.

3. These co-ordinates are evaluated by time dilation and relative velocities. These conform to the Lorentz transformations as well.
4. Calculations for co-ordinates of $E_1$ in the inertial frame co-moving with clock D: The inertial frames A and D synchronized their clocks A and D when the two clocks were at the same space-time point, event $E_2$ as x = x' = 0 and t = t' = 0.1. The Lorentz transformation between the two inertial frames can be written as

$$x' = [x - v(t - 0.1)] * \gamma$$

$$t' - 0.1 = \left[(t - 0.1) - \frac{vx}{c^2}\right] * \gamma$$

where $\gamma = \dfrac{1}{\sqrt{1 - v^2 / c^2}}$ and the primed co-ordinates are associated with the inertial frame co-moving with D.

With the normalized coordinates that we are using, $c = 1$. Further $v = 0.8$ according to the problem definition.

Substituting these values in the above equations and taking x = 0, t = 0 at event $E_1$, we get the event coordinates for $E_1$ in inertial frame co-moving with D as

x′ = 0.08*1.66667 = 0.133333 and

t′ = 0. 1 + (-0.1)*1.66667 = – 0.066667

The fact that three clocks in relative uniform motion could not be synchronized by the above process means that the clocks did not run equivalently. In this thought experiment we do not use spatially separated and synchronized clocks. Only the relativistic explanation uses the inertial frames co-moving with the point clocks and the spatially separated and synchronized clocks (synchronized as per the Einstein convention) in the respective inertial frames. The thought experiment as conceived has only three point clocks. Therefore, the issue of distant clock synchronization is eliminated. The three clocks are in uniform motion and do not experience any acceleration during the course of the thought experiment. The three point clocks which meet in pairs of three times (C (3, 2) = 3, same as 3 choose 2) should have been able to synchronize themselves if all of them were running equivalently. The fact that this did not happen shows that the first postulate of SRT is violated and a unique isotropic inertial frame exists.

It is to be noted that all the observations at events $E_1$, $E_2$ and $E_3$ are made by pairs of clocks that are present together at that particular event (here and now space-time point) and we do not require light rays or distant clock synchronization to record these comparisons.

The above thought experiment involving clocks A, B and D is an attempt to synchronize the three clocks assuming they run equivalently. The fact that this attempt failed, clearly means that the three clocks did not run equivalently.

See Appendix – A for space-time diagrams of this thought experiment from all the three inertial frames. See Appendix – B for a mathematical derivation based on Lorentz

Transformations that shows that at event – E3, the clocks B and D are found asynchronised confirming that the clocks A, B and D do not run equivalently.

## 5. Discussion

In Physics, we talk about an observer or an object being at rest within an inertial frame. We are usually at rest with respect to earth; when we are traveling we are at rest with respect to the car and moving with respect to the earth etc. This is in the spatial context.
But no object or observer in any frame, inertial or otherwise, can be at rest temporally.
This is the fundamental difference between space and time. Space and time cannot be mixed. They are two different independent dimensions. One (a person or a material object) can be at rest spatially in a certain context, but one can never be at rest in a temporal sense.

In the context of a time scale applicable to cosmology, Rugh and Zinkernagel [12] discuss scale setting processes that define length of time. It is clear that over the length of the cosmic time scale of 0 to $4*10^{17}$ seconds (about 14 billion years), the processes that can define a time scale may take different versions. The current process that we often refer to is based on the unit of year which is the time taken for earth to complete one orbit around the sun. But the sun and earth did not exist for most of the duration of the conceivable cosmic history. Therefore, it is obvious that different processes are needed to define the time scales in different epochs of the cosmic history. However, in order to be meaningful these time scales have to form a comparable measure to provide continuity in the measurement of time.

On another note, Rugh and Zinkernagel [12] discuss temporal order of events in the cosmic history and in our immediate temporal neighborhood, in the context of difficulties in defining consistent time scale by scale setting processes over the cosmic period of billions of 'years'. Temporal order does not measure duration but only the before, after or simultaneous nature of two events, and does not require a time scale. In our opinion, a relativistic frame-dependent temporal order will render the temporal order irrelevant. There needs to be an absolute temporal order by which one can say that two events A and B are such that either A preceded B or B preceded A or they were simultaneous. If such a temporal order is frame dependent, it will be difficult to assign any meaning to it.

Although a measure of time and the definition of a unit of time could be taking different forms over the cosmological time scale from epoch to epoch, one can conceptualize comparable measures over the transitions. Without this a cosmological time scale will not be feasible. In the context of an epoch (a temporal neighborhood) in the cosmological scale, well-defined units for measurement of time exist such as the atomic transition frequencies and half-life of unstable particles in the context of the current epoch wherein we are observing natural phenomena. These time units form the basis of the measurement of time in the clocks associated with the thought experiment of Section 4, and we can conclude that clocks A, B, and D described in Section 4 did not run equivalently.

This means that time dilation (clocks of moving inertial frames ticking at slower rate) is real and therefore, inertial frames are not equivalent in the sense that the clocks in different inertial frames tick at different rates.

## 6. Conclusion

The assumption of equivalence of all inertial frames in the sense that all inertial frames are isotropic along with the result of the Michelson-Morley experiment that the round trip speed of light is a constant, leads to the space-time continuum and reciprocal time dilation.

The process of slow separation of clocks (to synchronize clocks within an inertial frame) as described in Section 2 elucidates that if all inertial frames are isotropic and clocks tick at the same rate in all inertial frames, then asynchronization between spatially separated clocks of two inertial frames will not happen.

One particular clock of K and another particular clock of K' meet only once. In Section 4 by using three clocks meeting in pairs we clearly establish that the three of them did not run equivalently. It is seen that clock B must run faster than clock A or clock D or both. In other words all the three clocks cannot be running equivalently and clock B cannot be the slowest. This shows that clocks in different inertial frames run at different rates, justifying observed time dilation but proving that all inertial frames are not equivalent. The inertial frame wherein clocks run fastest is the isotropic inertial frame.

The thought experiment described in Section 4 shows that if all clocks ticked equally, then time dilation will not be observed. In view of the experimental observations of time dilation, we can infer that moving clocks run slow and this means there is a unique isotropic inertial frame with respect to which other inertial frames are moving. However, if one alters the synchronization in an anisotropic inertial frame by definition to that prescribed by the Einstein-Poincare convention, then it may appear that the anisotropic inertial frame is the isotropic inertial frame; this symmetry is only an illusion as explained in Section 2.

From above discussion we can conclude that there exists a unique isotropic inertial frame and all other inertial frames only appear to be isotropic within the framework of the Einstein-Poincare synchronization convention. In reality, absolute simultaneity or universal synchronization is the synchronization achieved in the unique isotropic inertial frame by the Einstein-Poincare synchronization convention.

A universal synchronization maintains the round trip speed of light as constant in all inertial frames. It avoids all paradoxes of reciprocal time dilation and reversal of time order.
A synchronization convention setting the speed of light in every direction in every inertial frame as *c*, creates a four dimensional continuum negating the evolution of future from the present. It creates paradoxes such as reciprocal time dilation and reversal of time order of events.

The creation of the ubiquitous four dimensional space-time continuum just by the adoption of a convention to synchronize spatially separated clocks is unconvincing to many students and scholars. The negation of the dynamic evolution of future from present by a ‘convention’ has not gained full acceptance, especially when the other choice is available. In this paper we have shown by theoretical considerations and a thought experiment that a unique isotropic inertial frame exists. Although the apparent symmetry as described in Section 2 and the associated difficulties in identifying the unique isotropic inertial frame are recognized, the quest to identify/locate the unique isotropic frame is ongoing. Until this quest comes to a fruitful conclusion we need to recognize that such a unique isotropic inertial frame exists.

## References

[1] Bohm, D., 1996, The Special Theory of Relativity, Routledge.

[2] Einstein, A., 1961, Relativity: The Special and the General Theory (Three Rivers Press, New York), Translated by Robert W Lawson.

[3] Iyer, C., and Prabhu, G. M., 2013, Synchronizing equivalent clocks across inertial frames, Int. J of Engg. Res. & Sci & Tech 2 (2), p. 8-17 https://www.ijerst.com/download.php?file=ijerstadmin/upload/IJEETC_5ccbe5a5701d3.pdf&iid=22

[4] Iyer, C., and Prabhu, G. M., 2006, Reversal in the time order of interactive events: the collision of inclined rods, Eur. J. Phys. Vol 27 (4), p. 819-824 https://doi.org/10.1088/0143-0807/27/4/013 , arXiv: 0809.1721.

[5] Iyer, C., and Prabhu, G.M., 2013, Moving signals and their measured frequencies, Int. J. Engg. Res. & Sci & Tech 2 (3), p. 24-36 https://www.ijerst.com/ijerstadmin/upload/IJEETC_5ccbf7143596b.pdf ; arXiv: 1201.3205.

[6] Janis, A., 2018, Conventionality of Simultaneity, *The Stanford Encyclopedia of Philosophy* (Fall 2018 Edition), Edward N. Zalta (ed.), URL = <https://plato.stanford.edu/archives/fall2018/entries/spacetime-convensimul/>.

[7] Kak, Subhash, 2007, Moving Observers in an Isotropic Universe, Intl. J. of Theoretical Physics (46) p. 1424-1430. https://ui.adsabs.harvard.edu/abs/2007IJTP...46.1424K/abstract https://arxiv.org/pdf/physics/0605199

[8] Lorentz, H. A., 1899, Simplified theory of electrical and optical phenomena in moving systems, Proc. Acad. Sci. Amsterdam 1, p. 427-442.

[9] Lorentz, H. A., 1904, Electromagnetic phenomena in a system moving with any velocity smaller than that of light, Proc. Acad. Sci. Amsterdam 6, p. 809-831.

[10] Michelson, A., and Morley, E., 1887, On the relative motion of the earth and the luminiferous ether, Am. J. Sci. 34 (203), p. 333-345.

[11] Reichenbach, H., 1957, The Philosophy of Space and Time (Dover, New York). First published in German under the title *Philosophie der Raum-Zeit-Lehre* in Walter de Gruyter, Berlin, 1927.

[12] Rugh, S.E., and Zinkernagel, H., 2009, On the physical basis of cosmic time, Studies in History and Philosophy of Science Part B: Studies in History and Philosophy of Modern Physics 40 (1): p. 1-19 https://arxiv.org/abs/0805.1947

**Appendix A: Space-Time Diagrams**

Space-time diagrams of the thought experiment depicted in the three inertial frames co-moving with clocks A, B, and D.

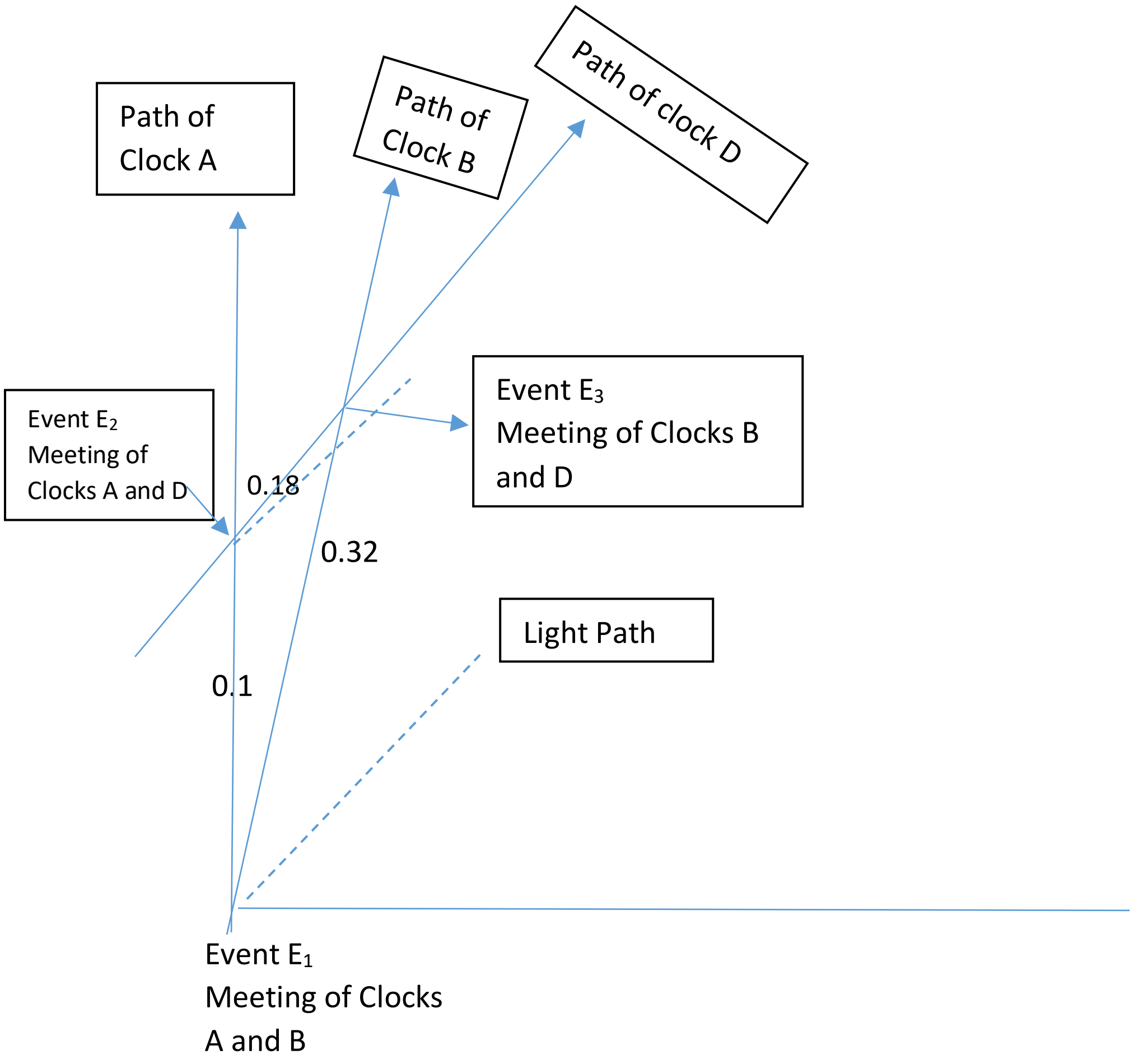

**Fig. 4. Space-time diagram viewed from inertial frame co-moving with clock A**

The co-ordinates of $E_1$, $E_2$, and $E_3$ for the inertial frame co-moving with clock A are given in column 2 of Table 2. $E_1$, $E_2$ and $E_3$ are the three events. Clocks A and B meet at $E_1$. Clocks A and D meet at $E_2$. Clocks B and D meet at $E_3$. In the space-time continuum the following statements can be made:

- Clock A travels along the line $E_1$ $E_2$
- Clock B travels along the line $E_1$ $E_3$
- Clock D travels along the line $E_2$ $E_3$
- In Euclidean geometry $E_1$ $E_2$ + $E_2$ $E_3$ is greater than $E_1$ $E_3$
- In space-time geometry $E_1$ $E_2$ + $E_2$ $E_3$ is less than $E_1$ $E_3$

All events $E_1$, $E_2$, $E_3$ exist forever, so to speak. So perception or consciousness has to travel. If $E_1$'s consciousness and $E_2$'s consciousness and $E_3$'s consciousness travel at the same rate, then in order to perceive these three events, $E_1E_2$ + $E_2$ $E_3$ should be equal to $E_1E_3$. But in space-time geometry $E_1E_2 + E_2E_3$ is less than $E_1E_3$.

Thus we can conclude that in order for B and D to meet at $E_3$, either A or D or both have to travel slower than B.

This can be compared with the familiar Euclidean geometry where in a triangle PQR, if A traces one side PQ, D traces another side QR and B traces side PR, and if A and B start together at P and D leaves Q at the same instant when A reaches Q, then in order that D meets B at R, either A or D or both have to travel faster than B.

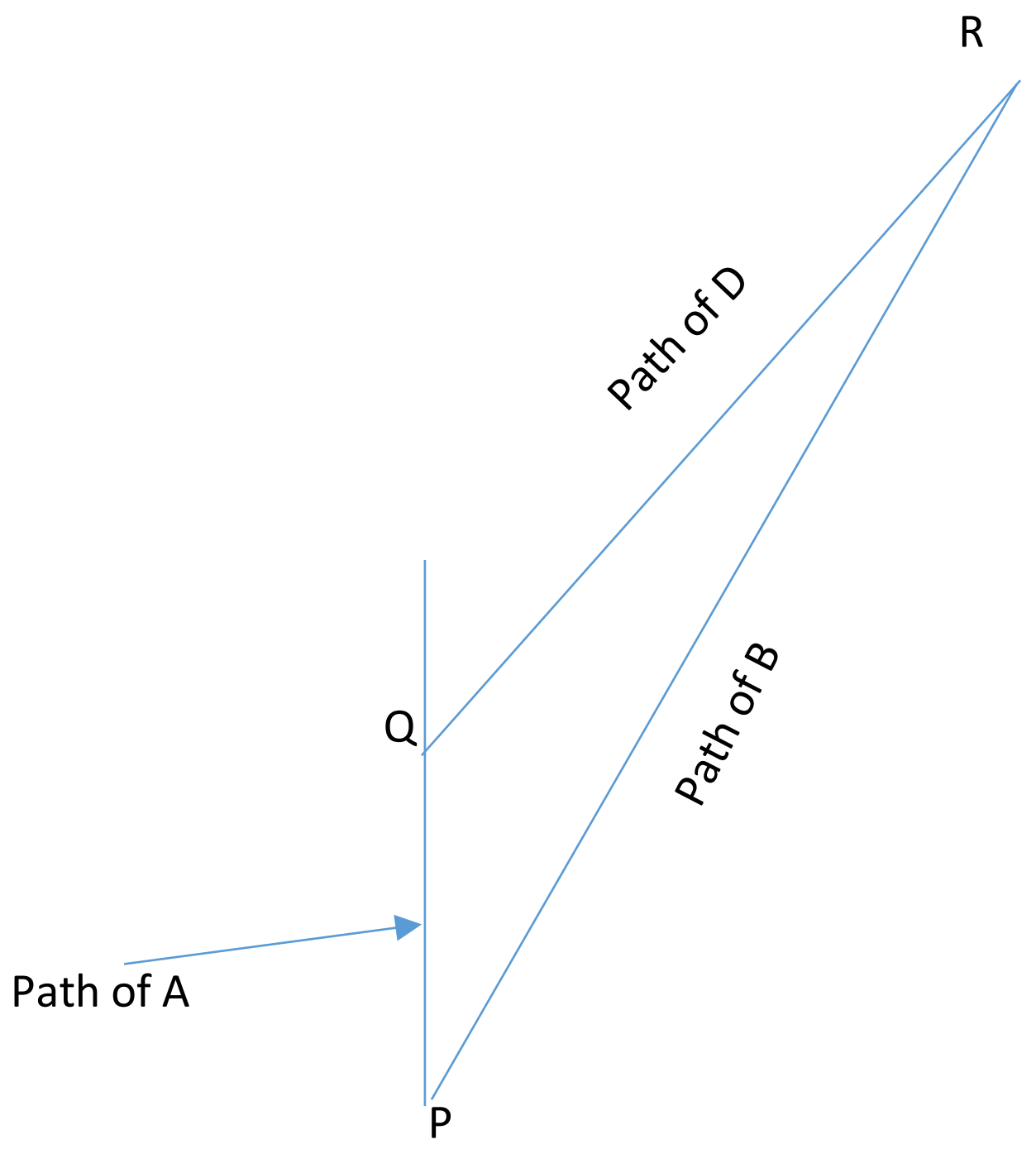

**Fig. 5. Analogous Euclidean representation**

Note: A and B start together at P. D leaves Q at the same instant that A reaches Q.

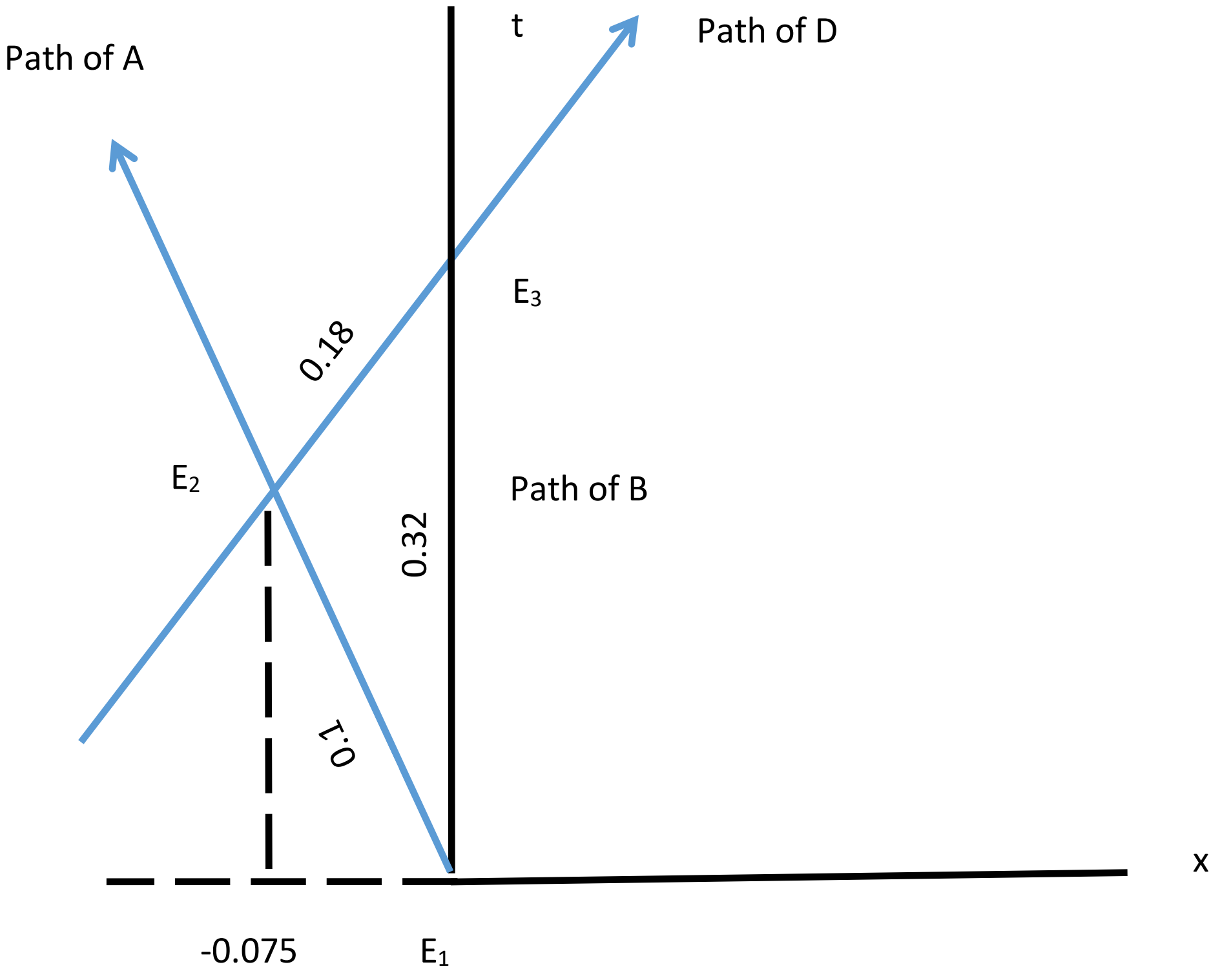

**Fig. 6. Space-time diagram viewed from inertial frame co-moving with clock B**

Co-ordinates of $E_1$, $E_2$, and $E_3$ for inertial frame co-moving with clock B are given in column 3 of Table 2.

The ‘distance’ between events is invariant and remains 0.10, 0.18, and 0.32 as indicated above, as observed in inertial frame A.

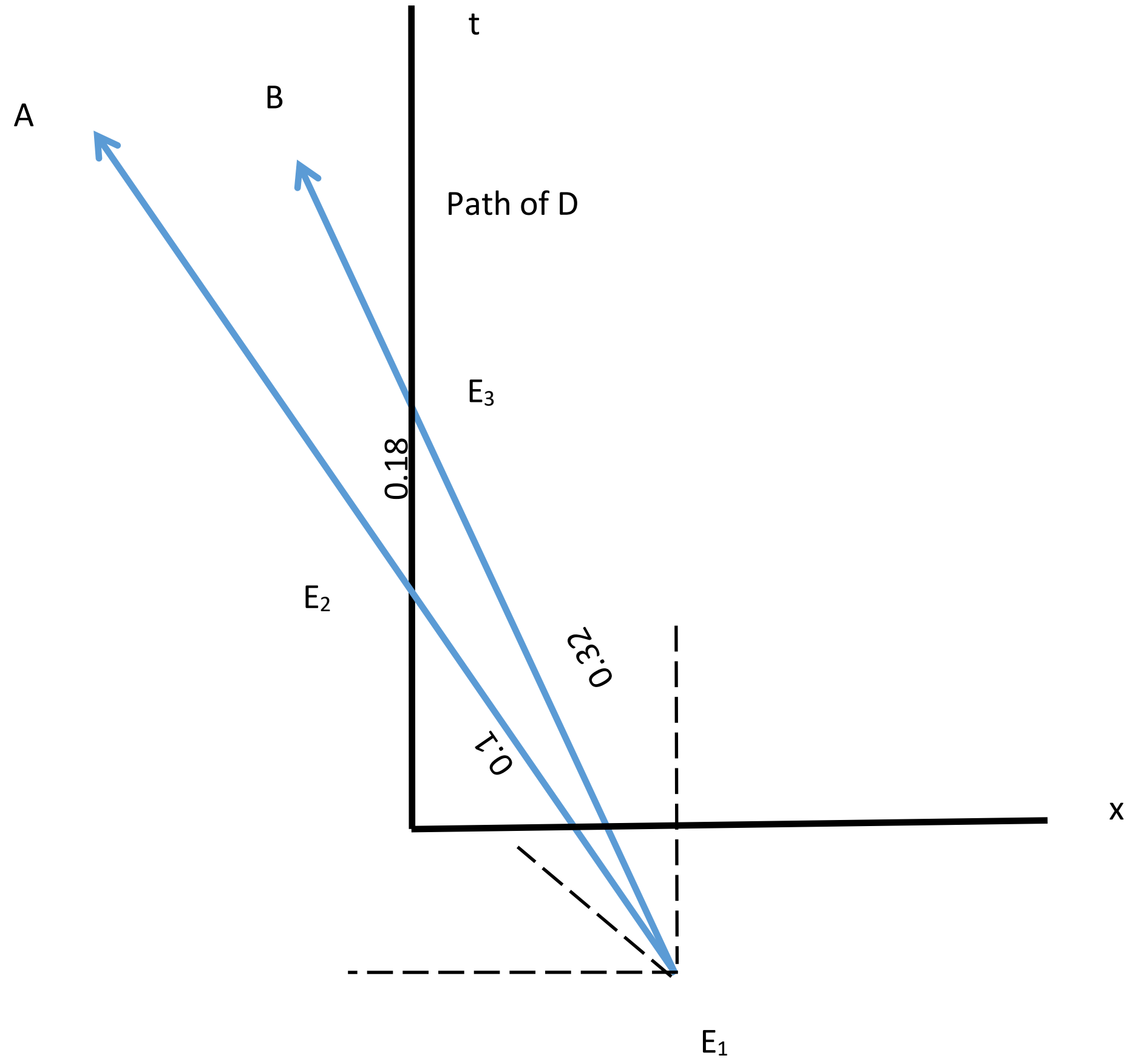

**Fig. 7. Space-time diagram viewed from inertial frame co-moving with clock D**

Co-ordinates of $E_1$, $E_2$, and $E_3$ for inertial frame co-moving with clock D are given in column 4 of Table 2.

The 'distance' between events is invariant and remains 0.10, 0.18, and 0.32 as indicated above, the same as observed in inertial frame A.

The distances will remain the same in any arbitrary inertial frame as well.

**Appendix B** - Derivation of the results by Lorentz Transformation (discrepancy at event -3)

Clock A (inertial frame K ) and clock B (inertial frame K' ) synchronise at x=0; t=0: x'=0; t'=0.

The speed between them is K' moving at 0.6c as observed by K. There synchronisation occurs when clock A and clock B pass each other.

Thus the LT between the frames K and K' is given by (using normalised coordinates where c=1)

$x' = (x-0.6t)*1.25$ -------- (1a)
$t' =(t-0.6x)*1.25$ ---------- (1b)
$x= (x'+0.6t')*1.25$ ------ (1c)
$t = (t'+0.6x')*1.25$ ----- (1d)

Clock A (inertial frame K ) and clock D (inertial frame K'' ) synchronise at x=0; t=0.1 ; x''=0; t''=0.1 when clock D passes clock A at t=0.1 The speed between them is K'' moving at 0.8c as observed by K. There synchronisation occurs when clock A and clock D pass each other. At that event, clock A reads 0.1 and clock D synchronised with clock A and sets the time as 0.1

Thus the LT between the frames K and K'' is given by (considering T=(t-0.1) and T"= (t"-0.1) as the hypothetical time coordinates of K and K", and recognising T=T"=0 when the origins of K and K" meet.)

$x'' = [x-0.8(t-0.1)]*(5/3)$ ------------- (2a)
$(t''-0.1) =[(t-0.1)-0.8x]*(5/3)$ --------- (2b)
$x= [x''+0.8(t''-0.1)]*(5/3)$ ------------ (2c)
$(t-0.1) = [(t''-0.1)+0.8x'']*(5/3)$----- (2d)

From the above, we can derive the transformation between K' and K'' as per the steps given below:

Substituting for x & t from (1c) and (1d) into equations of (2a) and (2b)
We get

$x'' = [(x'+0.6t')*1.25 – 0.8\{(t'+0.6x')*1.25 – 0.1\}](5/3)$ and -------------- (3)

$(t''-0.1) = [\{(t'+0.6x')*1.25\}-0.1 – 0.8* (x'+0.6t')*1.25] * (5/3)$ ---------- (4)

Simplifying (3) we get
$x'' = [x'-(5/13)t'](13/12) + (4/30)$ --------- (5)

or $x'' = (x'-0.3846 t')*1.08333 + 0.133333$ ------- (5)

Simplifying (4), we get

t’’ = {t’ – x’(5/13)}(13/12) - (2/30) ----- (6)

or t” = (t’ – 0.3846 x’) * 1.083333 – 0.0666667 ----- (6)

Above two equations give a direct transformation from K’ to K’’. The relative velocity between K’ and K” is equal to (5c/13) and corresponding value of gamma is (13/12). The inverse of the same can be easily obtained by algebraic rearrangement as

x’ = [x” + (5/13)t”](13/12) - (7/60) -------- (7)

or x’ = (x” + 0.3846 t”)* 1.083333 – 0.1166667 ------- (7)

t’ = [t”+ (5/13)x”] (13/12) + (1/60) ------ (8)

or t’ = (t” + 0.3846 x”) * 1.083333 + 0.01666667 ------- (8)

For the event when the origins of K’ and K” meet, we have x’ = x” = 0, Substituting x’=x” = 0 in equation (5), we get (5/12)t’ = (4/30) or **t’ = 0.32.** This is the time recorded by clock B when clocks B and D passed each other.

substituting x’=x”=0 in (7) we get (5/12) t” = 7/60 or **t” = 0.28.** This is the time recorded by clock D when clocks B and D passed each other.

In effect, what we have is clock A and B were synchronized at event 1 when they passed each other. Clocks A and D were synchronised when they passed each other at event 2. Later at event 3, when clocks B and D passed each other, it was observed that there was a time difference between them **(0.32 & 0.28** read by the clocks B and D respectively). This means, the clocks A, B and D did not tick identically during the duration of the thought experiment. This violates the equivalence of inertial frames in the sense identical behaviour of clocks in all the three inertial frames is not feasible under special theory of relativity, violating the first postulate of the theory, thus leading to a self-contradiction.

It is also possible to derive the relative speeds of the three inertial frames with respect to each other, when the observation at event E3 (namely the reading of the clocks 0.32 and 0.28) are a given. This can be done assuming the speeds as u and v and using the observation 0.32 and 0.28 at event E3.